\documentclass[a4paper, aps, prl, superscriptaddress, twocolumn]{revtex4-1}

\usepackage[utf8]{inputenc}
\usepackage[english]{babel}
\usepackage[dvipsnames]{xcolor}     
\usepackage{graphicx}
\usepackage{amsmath,amssymb}
\usepackage{amsfonts}
\usepackage{color}
\usepackage{mathtools}
\usepackage{braket}
\renewcommand{\eqref}[1]{(\ref{#1})}

\makeatletter
\renewcommand{\fnum@figure}{Fig. \thefigure}
\makeatother

\newcommand{\tensorarrow}[1]{\overset{\text{\tiny$\bm\leftrightarrow$}}{\bm{#1}}}

\usepackage{bm}
\renewcommand{\vec}{\ensuremath{\bm}}

\renewcommand{\Im}{\text{Im}}
\renewcommand{\Re}{\text{Re}}

\usepackage{hyperref}
\hypersetup{colorlinks,citecolor={blue},linkcolor={blue},urlcolor={blue}}

\renewcommand{\eqref}[1]{(\ref{#1})}

\usepackage{enumitem}

\begin{document}

\title{Optical polarization analogue in free electrons beams}
\date{\today}

\author{Hugo Lourenço-Martins}
\affiliation{University of Göttingen, IV. Physical Institute, Göttingen, Germany}
\author{Davy Gérard}
\affiliation{Light, nanomaterials, nanotechnologies (L2n), CNRS-ERL 7004, Université de Technologie de Troyes, Troyes 10000, France}
\author{Mathieu Kociak}
\affiliation{Laboratoire de Physique des Solides, Université Paris-Sud, CNRS-UMR 8502, Orsay 91405, France}

\begin{abstract}
Fast electrons spectromicroscopies enable to measure quantitatively the optical response of excitations with unrivaled spatial resolution. However, due to their inherently scalar nature, electron waves cannot access to polarization-related quantities. In spite of promising attempts based on the conversion of concepts originating from singular optics (such as vortex beams), the definition of an optical polarization analogue for fast electrons has remained a dead letter. Here, we establish such an analogue as the dipole transition vector of the electron between two well-chosen singular wave states. We show that electron energy-loss spectroscopy (EELS) allows a direct measurement of the \textit{polarized} electromagnetic local density of states. In particular, in the case of circular polarization, it measures directly the local optical spin density. This work establishes EELS as a quantitative technique to tackle fundamental issues in  nano-optics, such as super-chirality, the local polarization of dark excitations or polarization singularities at the nanoscale.
\end{abstract}

\maketitle

Classical wave optics and quantum mechanics share strong similarities rooted in the underlying - Helmholtz or Schrödinger - wave equations \cite{akkermans2004}. This very close resemblance is e.g. directly translated in the great amount of optical-electronic analogue phenomena, from the much celebrated Young-Feynman double-slits experiment \cite{Tonomura1989,Bach2013} to more exotic yet fascinating examples such as corrals \cite{Crommie1993,ColasdesFrancs2001} or Anderson localization \cite{Anderson1958,Wiersma1997}.\\
This mesmerizing analogy initiated a long standing and fruitful dialogue between these two fields. A famous example is the development of transmission electron microscopy (TEM), strongly inspired by optical concepts  \cite{Rose2013,Scherzer1949,Haider1998}. Conversely, electron microscopy also influenced its photonic counterpart through the discovery of holography \cite{Gabor1948}. This mutual influence culminated a decade ago when electron vortices have been predicted \cite{Bliokh2007} and measured \cite{Uchida2010,Verbeeck2010,Mcmorran2011} - these exotic beams constituting a canonical example of a generic wave phenomenon \cite{Nye1974} first observed with light \cite{Allen1992}.\\

This analogy was more recently extended to the inelastic interaction of light or electrons with matter, see e.g. \cite{DeGarciaAbajo2009a}. In particular, electron energy-loss spectroscopy (EELS) and light extinction spectroscopy (LES) gives extremely similar results when interacting with optical media \cite{Losquin2015}. Also, the spatial and spectral variations of the  EEL intensity can be described using the nano-optical concept of electromagnetic local density of states (EMLDOS) \cite{GarciadeAbajo2008}, and therefore gives access to bright as well as to dark modes \cite{GarciadeAbajo2008,Losquin2015,Losquin2015b}.
Nevertheless, EELS in an electron microscope is seriously hindered by its well-known inability to measure the polarization of photonic excitations, which is rooted in the scalar character of the Schrödinger equation. Now, the importance of polarization effects at the nanoscale is not to be demonstrated, and developing a polarized EELS (pEELS) could potentially shine light on sometimes controversial \cite{Kim2012} nanoscale polarization effects such as the super-chirality \cite{Collins2017,Tullius2015,Schaeferling2012}, i.e the local enhancement of circular dichroism beyond what is possible with a circularly polarized plane wave. 


 Recent advances have shown the potential of phase-shaped free electron beams, to reproduce optical polarization in EELS experiments. Indeed, the visionary work of Asenjo-Garcia and Garc\'ia de Abajo pointed to the use of vortex beams to mimick circular polarization \cite{Asenjo-Garcia2014}. Later,  Guzzinati et al. \cite{Guzzinati2017}  used $\pi$-beams - i.e. singular electron beams with a $\pi$ phase jump in the plane perpendicular to the electron propagation direction -  to emulate an optical polarization dependent experiment in EELS.  The selection-rule based approach developed in these works is similar in essence, although based on different physical assumptions \cite{Lourenco-Martins2020a}, to that developed for describing dichroic signals in the so-called core-loss EELS \cite{Schattschneider2014}. All together, these pioneering works, as well as the phenomenological work of Ugarte et Ducati \cite{Ugarte2016} and the numerical investigation of Zanfrognini and collaborators \cite{Zanfrognini2019} gave important hints on the relation between the symmetry of free electron beams and optical polarization. Unfortunately, they did not relate the EELS probabilities to any universal macroscopic or nanoscopic photonic observable. Additionally, it remained unclear what physical vectorial quantity for free electrons shall be used as an analogue to optical polarisation.

In this paper, we rigorously define an optical polarization analogue (OPA) for fast electrons as a vector equal to the transition dipole between two phase-shaped states. We then investigate the case where the beam waist of the electron beam $w_0$ is larger or comparable to the typical variation length of the probed nano-optical field $L$. We then demonstrate that the polarized EELS and the linear/circular optical extinction cross-sections can be directly connected, provided that incoming and outgoing electron states are properly defined. Particularly, we show the perfect analogy of the role of linear polarization dephasing upon wave propagation in the observation of circular dichroism in pEELS and LES.  In the case of nanoscale electron beams ($w_0 \ll L$) we  show  that pEELS measures the EMLDOS polarized linearly or circularly in the direction perpendicular to the electron propagation axis. This result sharply contrasts with conventional EELS experiments, which only access the component of the EMLDOS oriented along the beam propagation axis. Additionally, we demonstrate that the circular dichroism in pEELS is proportional to the local density of spin of the nano-optical field. \\

In the following, all the calculations make use of the quasistatic approximation. The electric field propagator  $\tensorarrow{G}(\vec{r},\vec{r}',\omega)$ and the electrostatic potential propagator  $W(\vec{r},\vec{r}',\omega)$ at points $\vec{r}$ and $\vec{r}'$ and frequency $\omega$ are therefore related through $4\pi \omega^2 \tensorarrow{G}(\vec{r},\vec{r}',\omega)=\vec{\nabla} \vec{\nabla}'W(\vec{r},\vec{r}',\omega)$. We use a modal decomposition  \cite{Ouyang1989,Boudarham2012} for performing simulations within and  \texttt{MNPBEM} \cite{Hohenester2014}, see SI for details. We describe the fast electron beam within the paraxial and non-recoil approximations (see SI) where the wavefunction $\psi(\vec{r})\propto \Psi(\vec{R})e^{ik_z z}$, $k_z$ being the wavevector along the propagation axis $z$ and $\vec{r}=(\vec{R},z)$.   The EELS probability related to a transition from the initial $\psi_i(\vec{r})$ and final $\psi_f(\vec{r})$ electron states with an energy-loss $\hbar \omega$ thus reads  \cite{GarciadeAbajo2010,Asenjo-Garcia2014,Guzzinati2017,Zanfrognini2019}:

\begin{equation}
\begin{split}
\Gamma_{if}(\omega) =& \dfrac{2 e^2}{h v^2} \iint d\vec{r}\,d\vec{r}'\; \Im\{-W(\vec{r},\vec{r}',\omega)\} \\ 
&\times \Psi_f^*(\vec{R})\Psi_i(\vec{R}) \Psi_f(\vec{R}')\Psi^*_i(\vec{R}') e^{-iq_z (z-z')}
\end{split}
\label{JGAequation}
\end{equation}

\noindent where $h$ is the Planck constant, $q_z=\omega/v$ denotes the longitudinal momentum transfer and $v\approx c/2$ is the electron speed in the TEM.

\begin{figure}[tbhp]
    \includegraphics[width=\columnwidth]{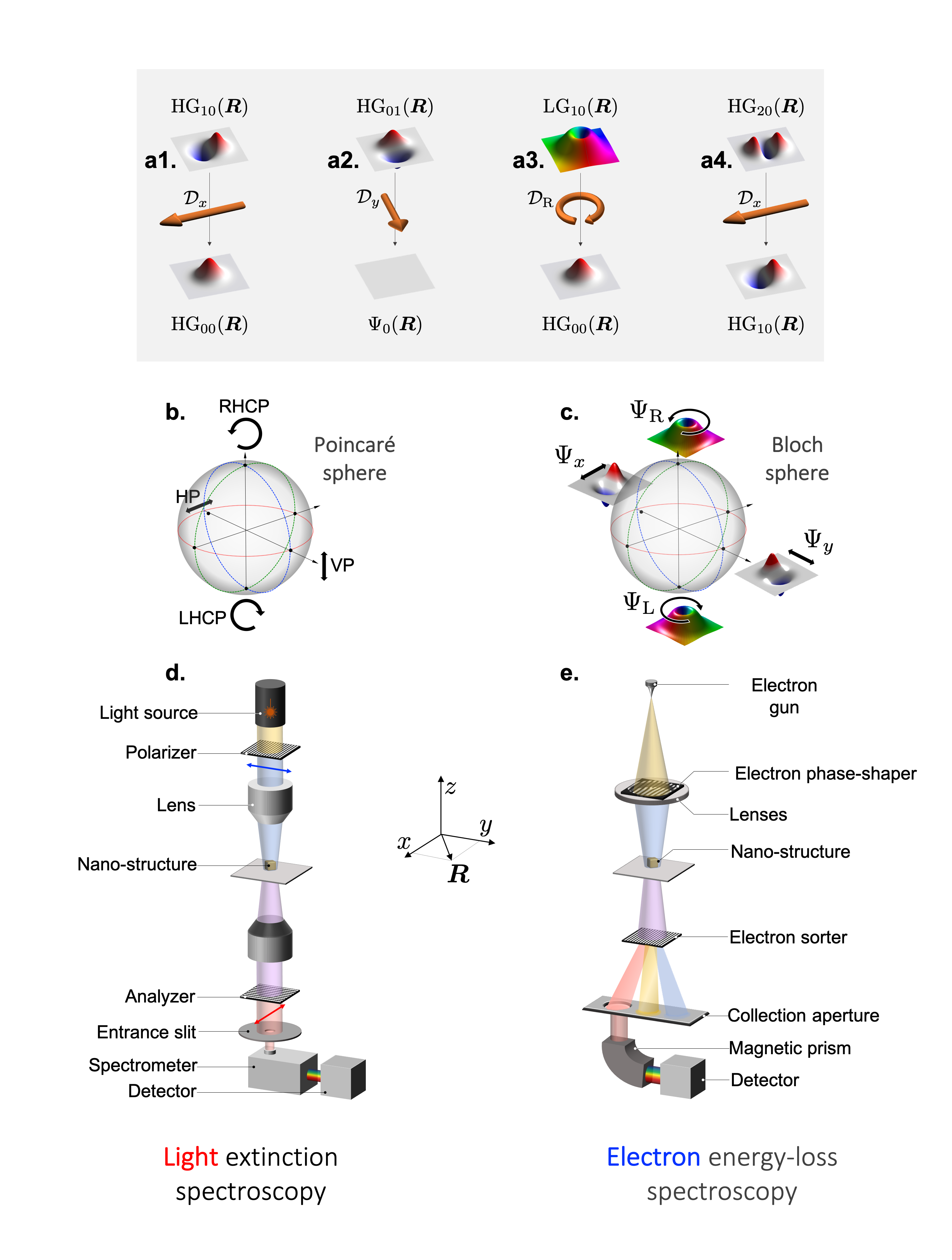}
    \caption{\textbf{Comparison of  polarized LES (left column) and non-spatially resolved polarized EELS (right column) experiments -} \textbf{a1.-a4.} Examples of four different transitions carrying a non-zero transition dipole moment. The relation between linear and circular polarizations (resp. $HG_{0|1,1|0}$ and $LG_{\pm 1,0)}$ wavefunctions) are represented on \textbf{b.} the Poincaré sphere (\textbf{c.} the Bloch sphere). The LES and pEELS experimental setups are represented respectively on figure \textbf{d.} and \textbf{e.}}
    \label{fig:Figure1}
\end{figure}

We compare LES and  pEELS experimental setups on Fig.~\ref{fig:Figure1}. 
pEELS requires  transitions between specific states to happen, as sketched on Fig.~\ref{fig:Figure1}(a1-a4). Without loss of generality, $\Psi_i$ is  supposed to be an idealized vortex beam, i.e, a  Laguerre-Gauss (LG$_{\pm 1,0)}$) \cite{Schattschneider2012} or a $\pi$-beam (Hermite-Gauss state HG$_{0|1,1|0}$)   \cite{Guzzinati2017}. $\Psi_{f}$ should be assumed to be a LG$_{0,0}$ (or equivalently an HG$_{0,0}$) state.  However, practical realizations of such experiments would require using a state sorter (see Fig. \ref{fig:Figure1}(d)), which is still subject to intense experimental research \cite{Grillo2017,Pozzi2020}. Therefore, the preferred realization is a very small spectrometer entrance aperture, i.e., a planewave $\Psi_0$ as a final state, which however share the same symmetry with the LG$_{0,0}$  and HG$_{0,0}$ states. A quantitative evaluation of the maximal size of the aperture possible before this approximation fails is given in the SI.    

 First, one needs to answer the question "\emph{how a scalar field (electron wavefunction) can project and measure a vectorial quantity (electromagnetic field)}"? \\
 For a LES experiment (see Fig.~\ref{fig:Figure1}(c)), the polarization effect directly follows from the fact that an external electromagnetic vectorial field, with e.g. an electrical field $\vec{E_u}$ ($u$ being either $x/y$ or $R/L$) interacts with another vectorial field, i.e, the polarization density in the medium.  In the electronic case, it seems really tempting to naively attribute the role of the external field to the transverse wave function $\Psi$. However, when drawing the analogy between the optical and electron cases, we must keep in mind a fundamental difference. In optical experiments, the probing wave (light) is directly involved in the process, while in electronic experiments, it is inelastically scattered \textit{via} the exchange of a photon. Therefore, in the later case, the symmetry of the \textit{transition}, and not of the wave itself as to be taken into account \cite{Asenjo-Garcia2014,Guzzinati2017}. In order to quantitatively understand this point, starting from equation \eqref{JGAequation}, one can write the transition from any first order HLG state to $\Psi_0$ as (see SI):
 \begin{equation}
\begin{split}
\Gamma_{\vec{u}} (\vec{R}_0,\omega) = \frac{4 q_z^2}{\hbar} \iint d\vec{r} \, d\vec{r}' \, \Im\{-\vec{j}_{\vec{u}}(\vec{r}).\tensorarrow{G}(\vec{r},\vec{r}',\omega).\vec{j}_{\vec{u}}(\vec{r}')\}
\label{EQ:general_OPA}
\end{split}
\end{equation}

\noindent where we have defined an effective transition current as:

\begin{equation}
\vec{j}_{\vec{u}}(\vec{r})=\vec{\mathcal{D}}_{\vec{u}}(\vec{R}) \, e^{iq_z z}     
\end{equation}

\noindent One can notice that this current has the form of an optical plane wave where the wave vector $q_z=\omega/v$ is the transferred momentum and the spatially dependent \emph{optical polarization analogue} of direction $\vec{u}$ is defined as: 
\begin{equation}
\vec{\mathcal{D}}_{\vec{u}}(\vec{R})=\braket{\Psi_0 \vert \hat{\vec{d}} \vert \Psi} f_{\vec{R}_0,w_0}(\vec{R})
\end{equation}
 
\noindent where $f_{\vec{R}_0,w_0}$ denotes a Gaussian function of center $\vec{R}_0$ and $w_0$ corresponding respectively to the impact point and width of the electron beam. We moreover introduced $\hat{\vec{d}}$ the \emph{transverse} transition dipole moment operator for the fast electron between any first order HLG state of the Bloch sphere $\Psi$ (see Fig.~\ref{fig:Figure1}(c)) to the plane wave $\Psi_0$. It simplifies to (see SI):
\begin{equation}
\braket{\Psi_0 \vert \hat{\vec{d}} \vert \Psi}= e w_0^2 \sqrt{2\pi} \vec{u}
\label{EQ:mapping_bloch_poincare}
\end{equation}
\noindent showing that $\vec{\mathcal{D}}_{\vec{u}}(\vec{R})$ is colinear to $\vec{u}$.

Expression \eqref{EQ:general_OPA} is now a scalar product as for the optical case. $\vec{\mathcal{D}_u}$  is a \textit{transition} dipole moment that defines a polarization analogue for EELS. Remarkably, in the case of a transition between $\ket{\Psi}$ to $\ket{\Psi_0}$, $\vec{u}$ corresponds to the point of the Poincaré sphere (see Fig.~\ref{fig:Figure1}(b)) located at the same coordinate as $\Psi$ on the Bloch sphere, giving an intuitive mapping between both spheres. However, more generally, this is the transition between the initial and the final state that gives rise to the vectorial form of the electron polarization analogue. Therefore, the mapping is less intuitive when considering   arbitrary transitions between HG$_{n,m}$ and HG$_{n',m'}$ respecting the selection rules $n'= n \pm 1$ or $m'=m \pm 1$ (linear case) or with LG$_{l,m}$ and LG$_{l',m}$ $l'=l\pm1$ (circular case) (see SI), i.e. in cases where the phase structure of the initial and final states have no obvious dipolar symmetry. We note that we haven't made any assumption on the energy transferred to the target, so that the above description is perfectly valid for probing any solid-state excitations, from phonons to core-loss excitation. 

 In the following, we will determine the optical observables measured with pEELS.  We first consider the broad illumination limit ($w_0 \gtrsim L$) in which light and electron beams can be compared as depicted in Fig.~\ref{fig:Figure1}. The extinction cross-section for a plane wave polarized along the direction $\vec{u}$ reads (see Fig.~\ref{fig:Figure1}(c) and SI):
 
\begin{equation}
\sigma_{\vec{u}}(\omega)=4\pi k_z^2  \iint d\vec{R} \, d\vec{R}' \;
\Im\left\{ \alpha_{uu}(\vec{R},\vec{R}',k_z,-k_z,\omega)\right\} 
\label{eq:broad_illumination_LES}
\end{equation}

\noindent where $\tensorarrow{\alpha}$ is the polarizability tensor and $k_z=\omega/c$ the wavevector of a plane wave propagating along $z$. This has to be compared to the electronic case where the  pEELS probability for a transition dipole $\vec{\mathcal{D}}_u$ can be deduced from \eqref{EQ:general_OPA} and reads:
\begin{equation}
\begin{split}
\Gamma_{\vec{u}}(\vec{R}_0, \omega)= 2 g q_z^2 \,\iint d\vec{R} \, d\vec{R}' \; f_{\vec{R}_0, w_0}(\vec{R}) \, f_{\vec{R}_0, w_0}(\vec{R}')\\ 
\Im\left\{- G_{\vec{u}\vec{u}}(\vec{R},\vec{R}',q_z,-q_z,\omega) \right\} 
\end{split}
\label{eq:broad_illumination_eels}
\end{equation}

\noindent where we have normalized the probability to the beam width (see SI) and defined $g=2e^2/h$ as the conductance quantum. Note that the polarizabilities and the Green dyadics are simply connected through a Dyson equation \cite{ColasdesFrancs2001}, which reduces to a simple proportionality relation (in the sense of tensors) for a dipole in vacuum.

\begin{center}
\begin{figure*}[tbhp]
    \includegraphics[width= 0.9\textwidth]{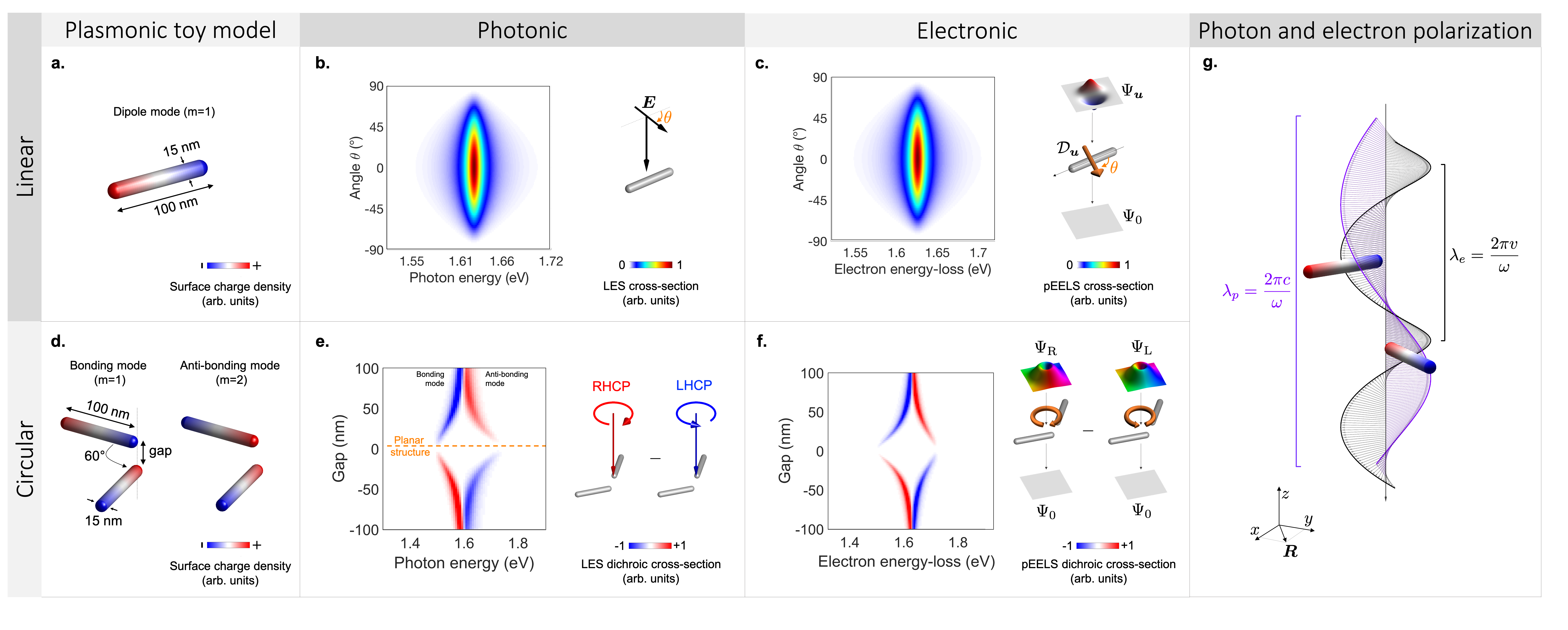}
    \caption{\textbf{Comparison of linear (top) and circulary (bottom) polarized LES and non-spatially resolved pEELS experiments on simple plasmonic nano-structures -}  \textbf{a.} Silver antenna sustaining a dipolar plasmon mode. Malus law measured on the antenna with \textbf{b.} light (LES) and \textbf{c.} electronic (pEELS) excitations. In this case, the electron beam is centered in the middle of the antenna. \textbf{d.} The simplest three-dimensional optically active plasmonic structure is built by combining two antennas similar to the one of \textbf{a}. These two antennas form an angle 60$^o$ and are offset along $z$ by a variable distance denoted as the gap. The activity increases with decreasing gap, in the same manner for \textbf{e.} optical and \textbf{f.} electronic measurements. In this case, the electron beam is centered on the tips of the two antennas. \textbf{g.} Schematics showing the propagation of a planewave of wavelength $\lambda_p$ (purple line) and of an effective electron transition current of wavelength $\lambda_e$ (black line) along a BKS nano-structure.}
    \label{fig:Figure2}
\end{figure*}
\end{center}

 The almost perfect resemblance between equations \eqref{eq:broad_illumination_LES} and \eqref{eq:broad_illumination_eels} shows that  pEELS is the counterpart of polarized optical extinction experiment, in the same vein as the correspondence between regular EELS and unpolarized extinction \cite{Losquin2015}. This is further exemplified for the linear polarization case in Fig.~\ref{fig:Figure2}(a-c), where the Malus law for the dipolar mode of a silver nanoantenna  is fully retrieved in both the photonic and electronic cases (Fig.~\ref{fig:Figure2}(b) and (c)), enforcing again the analogy between the optical polarization vector and $\vec{\mathcal{D}_u}$. Nevertheless, contrary to the optical case, the pEELS probability depends on the precise positioning of the beam through its dependence upon $\vec{R}_0$. 

Also of interest are the photonic circular dichroic extinction defined by $\mathcal{C}^\text{o}=\sigma_{L} - \sigma_{R} \propto \iint d\vec{R} \, d\vec{R}' \; \Re\left\{ \alpha_{xy} - \alpha_{yx}\right\}$
and the EELS one defined as $\mathcal{C}^e(\vec{R}_0,\omega)=\Gamma_R -\Gamma_L \propto \iint d\vec{R}\, d\vec{R}'\;  \Re\left\{  G_{yx}-G_{xy}\right\} f(\vec{R})f(\vec{R}')$ (see SI for the full formula). Clearly, $\mathcal{C}^\text{o}$ and $\mathcal{C}^e$ are also analogue.

This is exemplified when comparing circularly polarized LES and EELS for the simplest chiral plasmonic structure, the so-called Born-Kuhn model system (BKS, \cite{Yin2013a}). The BKS is built from two antennas  sustaining dipole resonances offset along the $z$ direction and rotated to each other (see Fig.~\ref{fig:Figure2}(d)). The gap and the angle fix the effective dephasing between the two subsequent interactions between the probe and each antenna. Both optic and electronic situations are almost identical with \emph{(i)} an increase of the dichroism visibility as the gap decrease and \emph{(ii)} an inversion of the dichroism when the sign of the gap flip - in perfect agreement with optical experiments reported in the litterature \cite{Lu2014}. The  strong  circular dichroic signals stem from the exact same physical ground. In both cases, the circular polarisation vector can be decomposed as a sum of linear polarization vectors $u_\text{R}=u_x\pm i u_y$. The linear polarization rotates upon propagation, exciting for the e.g $R$ polarization in phase (antiphase) the two dipolar charge distributions of the bounding (antibounding)  mode (Fig.~\ref{fig:Figure2}(g)). The only difference is the rotation speed of the linear polarization, being related to the light wavelength $c/\omega$ and the wavelength of the electromagnetic field following the electrons $v/\omega$. Since the electron speed can be changed at will, this makes EELS a quite tunable tool for the investigation of chiral structures, as already suggested for photon induced near-field microscopy (PINEM) \cite{Harvey2020}.

\begin{center}
\begin{figure}[bthp]
    \includegraphics[width= 0.9\columnwidth]{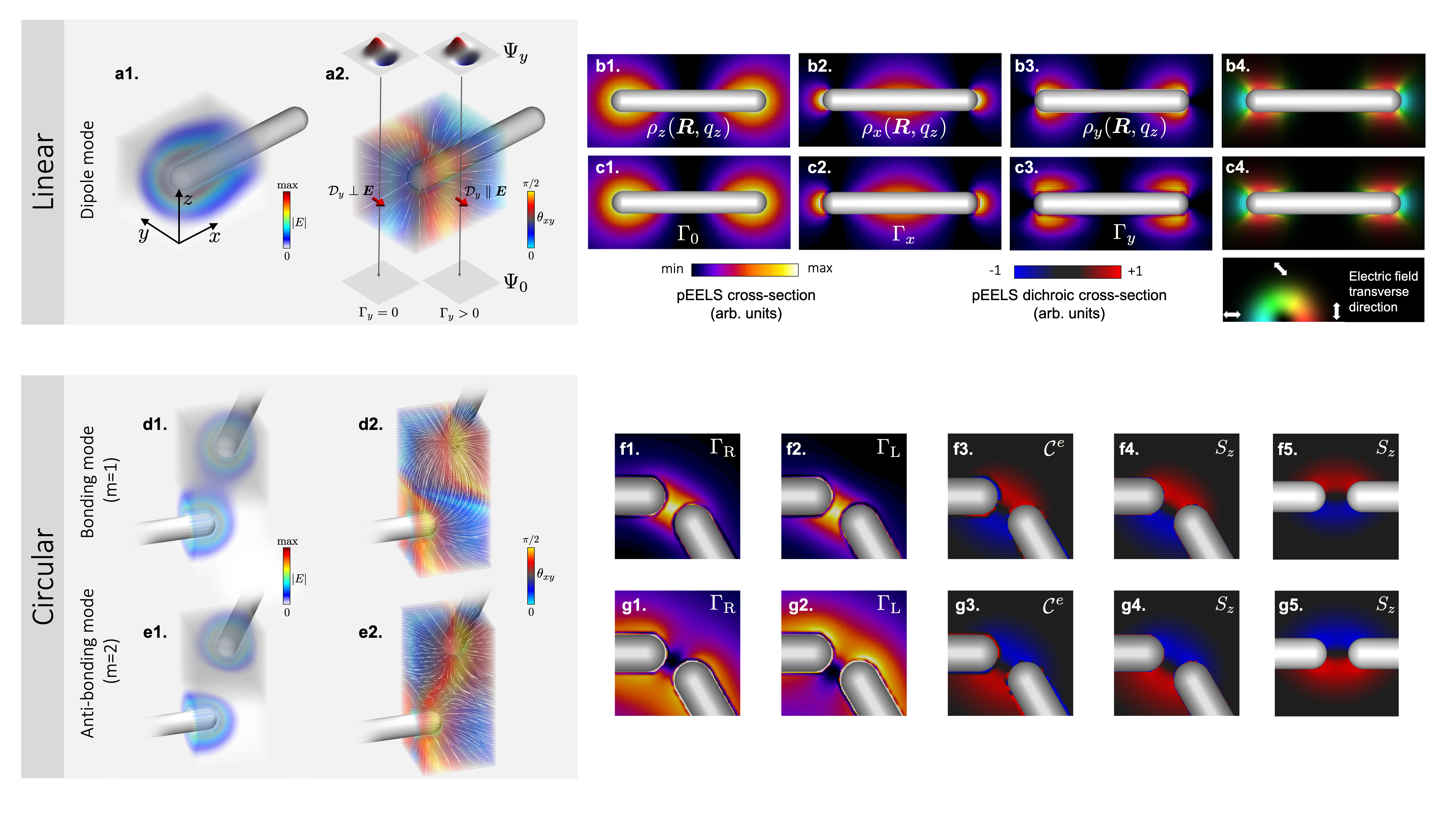}
    \caption{\textbf{Spatially resolved pEELS -} \underline{Top:} Spatially-resolved linear pEELS of the dipolar mode of a silver nanoantennas. \textbf{a1.} Three dimensional map of the electric field magnitude of the dipole mode (see Fig~\ref{fig:Figure2}(a)) around one tip of the rod. \textbf{a2.} Three dimensional map of the transverse direction $\theta_{xy}$ of the electric field (see main text). The field lines are represented in white. The dipole moment $\mathcal{D}_y$ of the transition $\Psi_y \rightarrow \Psi_0$ is represented at two different positions of the electron beam. \textbf{b.} FT-EMLDOS calculated along the \textbf{b1.} $z$-axis, \textbf{b2.} $x$-axis and the \textbf{b3.} $y$-axis. \textbf{c.} pEELS maps simulated with a \textbf{c1.} $\Psi_{0}$, \textbf{c2.} $\Psi_{x}$ and \textbf{c3.} $\Psi_{y}$ wavefunction. \textbf{b4.-c4.} Electric field transverse direction reconstructed respectively from (b2,b3) and (c2,c3) and plotted using a domain coloring method. \underline{Bottom:} Spatially-resolved circular dichroic pEELS signal for the BKS considered on Fig~\ref{fig:Figure2}(d) with a gap fixed at 25 nm. \textbf{d1-d2} Three dimensional map of the electric field magnitude and transverse direction $\theta_{xy}$ of the bonding mode. The field lines are represented in white. \textbf{e1-e2} Same maps but for the anti-bonding mode. pEELS maps of the bonding mode calculated for \textbf{f1.} a $\Psi_\text{R}$ and \textbf{f2.} a $\Psi_\text{L}$ wavefunction. \textbf{f3.} Dichroic pEELS map deduced from (f1-f2). \textbf{f4.} Map of the optical field vorticity along $z$ of the bonding mode. \textbf{g1-g4.} Same calculation as in (e1-e4) but for the anti-bonding mode. \textbf{f5, g5.} Same maps as (f4) and (g4) except that the BKS structure is aligned.}
     \label{fig:Figure3}
    \end{figure}
\end{center}

We now turn to the focused illumination limit ($w_0 \ll L$), see Fig.~\ref{fig:Figure3}(a,d). Using the fact that  $ \lim_{w\to 0}\Psi_x(x)=\delta'(x)$ for $\pi$ beams and  $\delta'(z)= \delta'(x) - i \delta'(y)$ for the $l=+1$  vortex beams (see SI), we  find  for a spatially resolved pEELS experiment between states $i$ and $f$ (see SI):
\begin{equation}
\Gamma_{\vec{u}}^{i,f}(\vec{R}_0,\omega)=\frac{2\pi q_z^2}{\hbar\omega} \; \vert\vec{d}_{i,f}\vert^2  \; \tilde{\rho}_{\vec{u}\vec{u}}(\vec{R}_0,q_z,\omega)
\label{EQ:pEELS_EMLDOS}
\end{equation}

\noindent where $\tilde{\rho}_{\vec{u}\vec{u}}(\vec{R}_0,q_z,\omega)$ denotes the Fourier transformed (along $z$) $\vec{u}$-polarized EMLDOS defined as \cite{GarciadeAbajo2008}:
\begin{equation}
\rho_{\vec{u}}(\vec{r},\omega)=-\frac{2\omega}{\pi} \Im\left\{\vec{u}^*.\tensorarrow{G}(\vec{r},\vec{r},\omega).\vec{u} \right\}    
\label{EQ:EMLDOS_definition}
\end{equation}

\noindent and $\vec{u}=\vec{d}_{if}/|\vec{d}_{if}|$ is the direction of the in-plane transition dipole $\vec{d}_{if}$.
We emphasize that, since $\vec{u}$ is any polarization of the Poincaré sphere, the latter equation extends the definition of the chiral EMLDOS of Pham et al. \cite{Pham2016,Pham2018}. Remarkably, one could notice that equation \eqref{EQ:pEELS_EMLDOS} is extremely similar to the Purcell formula \cite{Novotny2006} (see SI), with the electron transition dipole having the role of probe dipole. It is thus demonstrating that the transverse free electron state behaves analogously to a two-level system interacting with an nano-optical field, where the interaction time is encoded in the $z$-FT.\\

\noindent  To illustrate our findings, we simulated a spatially resolved pEELS experiment on the dipole mode of the same antenna as in Fig~\ref{fig:Figure2}(a). On Fig~\ref{fig:Figure3}(a1,a2), we respectively computed the three-dimentional map of the magnitude $\vert \vec{E} \vert$ and transverse direction $\theta_{xy}=\arctan(\vert E_y \vert/\vert E_x \vert)$ of the plasmonic electric field. As expected, one can  observe that the FT $z$-EMLDOS (Fig~\ref{fig:Figure3}(b1)) clearly reproduces the variations of the field magnitude (Fig~\ref{fig:Figure3}(a1)), while the FT $x$- and $y$-EMLDOS map the regions where the field is aligned along $x$ and $y$, respectively in blue and yellow on Fig~\ref{fig:Figure3}(a2). This simple correspondance comes from the fact that the transverse direction of the electric field does not strongly vary as a function of $z$, see Fig~\ref{fig:Figure3}(a2). As illustrated in Fig~\ref{fig:Figure3}(a2), we then simulated pEELS maps for $\Psi_{0} \rightarrow \Psi_{0}$ ($\vec{\mathcal{D}_u}=\vec{0}$), $\Psi_{x} \rightarrow \Psi_{0} $ $\vec{\mathcal{D}_u}=\vec{\mathcal{D}_x}$ and $\Psi_{y} \rightarrow \Psi_{0} $ ($\vec{\mathcal{D}_u}=\vec{\mathcal{D}_y}$) transitions respectively on Fig~\ref{fig:Figure2}(c1), (c2) and (c3). One can observe an almost perfect agreement between the pEMLDOS and the pEELS maps, which strongly supports our theoretical conclusions \eqref{EQ:pEELS_EMLDOS}. This result provides a clear and rigorous interpretation of the early local mapping of dipolar plasmons with a $\pi$-beam \cite{Guzzinati2017}. Additionally, from two different linear pEELS measurements, one can reconstruct the local polarization of optical fields (see Fig.~\ref{fig:Figure3}(b4-c4)), as otherwise measured by Krehl et al. with differential phase contrast imaging \cite{Krehl2018}. Our technique thus constitutes the ideal tool to resolve polarization singularities at the nano-scale. \cite{Burresi2009,Rotenberg2015}.

\onecolumngrid
\twocolumngrid 

\noindent Plugging \eqref{EQ:pEELS_EMLDOS} in $\mathcal{C}^e=\Gamma_\text{R}-\Gamma_\text{L}$, one can also rigorously define the spatially-resolved dichroic pEELS probability (see SI) as:
\begin{subequations}
\begin{align}
  \mathcal{C}^e  &= \dfrac{q_z^2 e^2}{ \hbar\omega } \; \big( \tilde{\rho}_{\text{R}}(\vec{R}_0,q_z,\omega) - \tilde{\rho}_{\text{L}}(\vec{R}_0,q_z,\omega)\big) \label{EQ:electron_dichroism1}\\
   &= 2 g q_z^2 \; \Re\left\{  G_{xy}(\vec{R}_0,q_z,\omega)-G_{yx}(\vec{R}_0,q_z,\omega) \right\}\label{EQ:electron_dichroism2}
\end{align}
\end{subequations}

\noindent where the dependence on $\vec{R}_0, ~\omega$ and $q_z$ has been omitted for brevity in the left-hand side. This equation allows us to formally define the optical dichroism at the nanoscale as the measure of the \emph{local} difference between the density of right- and left-handed optical states. Additionally, one can recognize in both equations \eqref{EQ:electron_dichroism1} and \eqref{EQ:electron_dichroism2} the definition of the $z$-component of the spin operator of the electromagnetic field applied on an plane wave of frequency $q_z$ \cite{Kim2012,Coles2012}:
\begin{equation}
\mathcal{C}^e(\vec{R}_0,q_z,\omega)=\dfrac{q_z^2 e^2}{\hbar^2\omega}S_z(\vec{R}_0,q_z,\omega)
\label{EQ:electron_dichroism3}
\end{equation}

\noindent The latter equation shows that $\mathcal{C}^e$ measures the $q_z$ component of the optical spin density along the direction of propagation of the electron beam which is itself proportional to the optical chirality flow \cite{Bliokh2011}. 
\noindent In order to clearly illustrate the physics at play here, we numerically investigated the nano-optical dichroism of the BKS introduced on Fig~\ref{fig:Figure2}(d) with a gap value fixed at 25 nm. On Fig~\ref{fig:Figure3}(d1,d2) and (e1,e2), we show the maps of the magnitude and transverse direction of the plasmonic electric fields respectively associated with the bonding and anti-bonding modes \footnote{Note that, for visibility, the gap has been increased to 50 nm on these plots}. Crucially, and as a signature of the chiral nature of the BKS, one can observe on Fig~\ref{fig:Figure3}(d2,e2) that  the transverse direction of the electric field rotates as a function of $z$, as the iso-direction regions form helices (in blue and yellow respectively on Fig~\ref{fig:Figure3}(d2) and (e2)). The local spin density $S_z(\vec{R}_0,q_z,\omega)$ is a direct measure of this property and quantifies in which direction the electric fields rotates along $z$ at $\vec{R}_0$ and for the spatial frequency $q_z$. This is illustrated, on Fig~\ref{fig:Figure2}(f1-f4) in the case of the bonding mode and on Fig~\ref{fig:Figure2}(g1-g4) for the anti-bonding mode. The perfect agreement observed between the maps of $\mathcal{C}^e$ and $S_z$ moreover corroborate our equations \eqref{EQ:electron_dichroism1}, \eqref{EQ:electron_dichroism2} and \eqref{EQ:electron_dichroism3}.\\
In order to give a more intuitive understanding, one can apply a modal decomposition to the spin operator (see SI and \cite{Boudarham2012}) and obtain $S_z\propto\Im\{E_m \times E^*_m\}_z$, $m$ being an integer indexing optical modes of the nano-structure. We here retrieve the Minkowski formula for the spin of optical fields \cite{Cameron2012b,Bliokh2013} which clearly show its connection with the local twist of the electric field.\\
\noindent Finally, one shall emphasize that these results shine a new light onto the microscopic origin of the extrinsic dichroism and its macroscopic expression. A nano-optical probe such as a phase-shaped electron beam measures  the local difference between the right and left EMLDOS. A broad beam measures the spatially integrated difference between left and right density of states. Consequently, a nano-structure can well be optically inactive  while having a non-zero density of spin (i.e. the nano-optical field exhibit a local twist), as illustrated on the non-chiral aligned BKS nanostructure (Fig. \ref{fig:Figure3}(f5-g5)). Equation \eqref{EQ:EMLDOS_definition} can be interpreted as the  Purcell factor for a \textit{chiral} molecules placed at point $R_0$ with the transition dipole of the tranverse electron wavefunction. Therefore, chiral pEELS can directly probe location of enhanced emission of chiral molecules even in globally achiral structures.

\section{Conclusion} 
We have shown that we can rigourously define an optical polarization analogue for the free electron beam. This leads to the possibility of introducing  polarized measurements in EELS. These are directly analogue to LES experiments in the case of broad beam illuminations. Spatially resolved pEELS map polarized EMLDOSs, and the dichroic circular pEELS probabilities are directly related to the density of electromagnetic spins. In particular, this permits a comprehensive description of the local polarization of both bright and dark optical excitations, while the otherwise highly sucessfull cathodoluminescence \cite{Gomez-Medina2008,Osorio2016} or PINEM \cite{Yurtsever2012,Harvey2020} polarized experiments are restricted to bright ones. Remarkably, through the mapping between Bloch and Poincaré spheres, our work establishes a Jones formalism for electrons. Thus, through the consideration of partial OPA, the development of a full polarimetric \cite{Osorio2016} pEELS is now at hand.
This study concentrates on the quasi-static limit, where magneto-electric effects are not taken into account. Extension to the relativistic case has already be described for circular pEELS of molecules \cite{Asenjo-Garcia2014}, but should be continued in a similar framework as the one developed here. Also, the formalism used here could be use as a basis to describe cathodoluminescence and PINEM experiments with phase-shape electrons.\\

\noindent We acknowledge J. Verbeeck for introduction to this field and thank J. Verbeeck, D. Ugarte, F. Houdellier and G. Guzzinati  for insightful discussions. H.L-M thanks Tyler R. Harvey for insightful discussions. This project has received funding from the European Union’s Horizon 2020 research and innovation programme under grant agreement No. 823717-ESTEEM3 and from the French state managed by the National Agency for Research under the programme of future investment EQUIPEX TEMPOS-CHROMATEM with the reference ANR-10-EQPX-50.

\bibliographystyle{apsrev4-1}

\begin{thebibliography}{55}%
\makeatletter
\providecommand \@ifxundefined [1]{%
 \@ifx{#1\undefined}
}%
\providecommand \@ifnum [1]{%
 \ifnum #1\expandafter \@firstoftwo
 \else \expandafter \@secondoftwo
 \fi
}%
\providecommand \@ifx [1]{%
 \ifx #1\expandafter \@firstoftwo
 \else \expandafter \@secondoftwo
 \fi
}%
\providecommand \natexlab [1]{#1}%
\providecommand \enquote  [1]{``#1''}%
\providecommand \bibnamefont  [1]{#1}%
\providecommand \bibfnamefont [1]{#1}%
\providecommand \citenamefont [1]{#1}%
\providecommand \href@noop [0]{\@secondoftwo}%
\providecommand \href [0]{\begingroup \@sanitize@url \@href}%
\providecommand \@href[1]{\@@startlink{#1}\@@href}%
\providecommand \@@href[1]{\endgroup#1\@@endlink}%
\providecommand \@sanitize@url [0]{\catcode `\\12\catcode `\$12\catcode
  `\&12\catcode `\#12\catcode `\^12\catcode `\_12\catcode `\%12\relax}%
\providecommand \@@startlink[1]{}%
\providecommand \@@endlink[0]{}%
\providecommand \url  [0]{\begingroup\@sanitize@url \@url }%
\providecommand \@url [1]{\endgroup\@href {#1}{\urlprefix }}%
\providecommand \urlprefix  [0]{URL }%
\providecommand \Eprint [0]{\href }%
\providecommand \doibase [0]{http://dx.doi.org/}%
\providecommand \selectlanguage [0]{\@gobble}%
\providecommand \bibinfo  [0]{\@secondoftwo}%
\providecommand \bibfield  [0]{\@secondoftwo}%
\providecommand \translation [1]{[#1]}%
\providecommand \BibitemOpen [0]{}%
\providecommand \bibitemStop [0]{}%
\providecommand \bibitemNoStop [0]{.\EOS\space}%
\providecommand \EOS [0]{\spacefactor3000\relax}%
\providecommand \BibitemShut  [1]{\csname bibitem#1\endcsname}%
\let\auto@bib@innerbib\@empty
\bibitem [{\citenamefont {Akkermans}\ and\ \citenamefont
  {Montanbaux}(2012)}]{akkermans2004}%
  \BibitemOpen
  \bibfield  {author} {\bibinfo {author} {\bibfnamefont {E.}~\bibnamefont
  {Akkermans}}\ and\ \bibinfo {author} {\bibfnamefont {G.}~\bibnamefont
  {Montanbaux}},\ }\href
  {http://books.google.com/books?id=GBCaauKNz5cC{\&}pgis=1} {\emph {\bibinfo
  {title} {{Physique m{\'{e}}soscopique des {\'{e}}lectrons et des
  photons}}}},\ Savoirs actuels\ (\bibinfo  {publisher} {EDP Sciences},\
  \bibinfo {year} {2012})\BibitemShut {NoStop}%
\bibitem [{\citenamefont {Tonomura}\ \emph {et~al.}(1989)\citenamefont
  {Tonomura}, \citenamefont {Endo}, \citenamefont {Matsuda}, \citenamefont
  {Kawasaki},\ and\ \citenamefont {Ezawa}}]{Tonomura1989}%
  \BibitemOpen
  \bibfield  {author} {\bibinfo {author} {\bibfnamefont {A.}~\bibnamefont
  {Tonomura}}, \bibinfo {author} {\bibfnamefont {J.}~\bibnamefont {Endo}},
  \bibinfo {author} {\bibfnamefont {T.}~\bibnamefont {Matsuda}}, \bibinfo
  {author} {\bibfnamefont {T.}~\bibnamefont {Kawasaki}}, \ and\ \bibinfo
  {author} {\bibfnamefont {H.}~\bibnamefont {Ezawa}},\ }\href {\doibase
  10.1119/1.16104} {\bibfield  {journal} {\bibinfo  {journal} {American Journal
  of Physics}\ }\textbf {\bibinfo {volume} {57}},\ \bibinfo {pages} {117}
  (\bibinfo {year} {1989})}\BibitemShut {NoStop}%
\bibitem [{\citenamefont {Bach}\ \emph {et~al.}(2013)\citenamefont {Bach},
  \citenamefont {Pope}, \citenamefont {Liou},\ and\ \citenamefont
  {Batelaan}}]{Bach2013}%
  \BibitemOpen
  \bibfield  {author} {\bibinfo {author} {\bibfnamefont {R.}~\bibnamefont
  {Bach}}, \bibinfo {author} {\bibfnamefont {D.}~\bibnamefont {Pope}}, \bibinfo
  {author} {\bibfnamefont {S.-H.}\ \bibnamefont {Liou}}, \ and\ \bibinfo
  {author} {\bibfnamefont {H.}~\bibnamefont {Batelaan}},\ }\href {\doibase
  10.1088/1367-2630/15/3/033018} {\bibfield  {journal} {\bibinfo  {journal}
  {New Journal of Physics}\ }\textbf {\bibinfo {volume} {15}},\ \bibinfo
  {pages} {033018} (\bibinfo {year} {2013})}\BibitemShut {NoStop}%
\bibitem [{\citenamefont {Crommie}\ \emph {et~al.}(1993)\citenamefont
  {Crommie}, \citenamefont {Lutz},\ and\ \citenamefont {Eigler}}]{Crommie1993}%
  \BibitemOpen
  \bibfield  {author} {\bibinfo {author} {\bibfnamefont {M.~F.}\ \bibnamefont
  {Crommie}}, \bibinfo {author} {\bibfnamefont {C.~P.}\ \bibnamefont {Lutz}}, \
  and\ \bibinfo {author} {\bibfnamefont {D.~M.}\ \bibnamefont {Eigler}},\
  }\href {\doibase 10.1126/science.262.5131.218} {\bibfield  {journal}
  {\bibinfo  {journal} {Science}\ }\textbf {\bibinfo {volume} {262}},\ \bibinfo
  {pages} {218} (\bibinfo {year} {1993})}\BibitemShut {NoStop}%
\bibitem [{\citenamefont {{Colas des Francs}}\ \emph
  {et~al.}(2001)\citenamefont {{Colas des Francs}}, \citenamefont {Girard},
  \citenamefont {Weeber}, \citenamefont {Chicane}, \citenamefont {David},
  \citenamefont {Dereux},\ and\ \citenamefont {Peyrade}}]{ColasdesFrancs2001}%
  \BibitemOpen
  \bibfield  {author} {\bibinfo {author} {\bibfnamefont {G.}~\bibnamefont
  {{Colas des Francs}}}, \bibinfo {author} {\bibfnamefont {C.}~\bibnamefont
  {Girard}}, \bibinfo {author} {\bibfnamefont {J.~C.}\ \bibnamefont {Weeber}},
  \bibinfo {author} {\bibfnamefont {C.}~\bibnamefont {Chicane}}, \bibinfo
  {author} {\bibfnamefont {T.}~\bibnamefont {David}}, \bibinfo {author}
  {\bibfnamefont {A.}~\bibnamefont {Dereux}}, \ and\ \bibinfo {author}
  {\bibfnamefont {D.}~\bibnamefont {Peyrade}},\ }\href {\doibase
  10.1103/PhysRevLett.86.4950} {\bibfield  {journal} {\bibinfo  {journal}
  {Physical Review Letters}\ }\textbf {\bibinfo {volume} {86}},\ \bibinfo
  {pages} {4950} (\bibinfo {year} {2001})}\BibitemShut {NoStop}%
\bibitem [{\citenamefont {Anderson}(1958)}]{Anderson1958}%
  \BibitemOpen
  \bibfield  {author} {\bibinfo {author} {\bibfnamefont {P.~W.}\ \bibnamefont
  {Anderson}},\ }\href {\doibase 10.1103/PhysRev.109.1492} {\bibfield
  {journal} {\bibinfo  {journal} {Physical Review}\ }\textbf {\bibinfo {volume}
  {109}},\ \bibinfo {pages} {1492} (\bibinfo {year} {1958})}\BibitemShut
  {NoStop}%
\bibitem [{\citenamefont {Wiersma}\ \emph {et~al.}(1997)\citenamefont
  {Wiersma}, \citenamefont {Bartolini}, \citenamefont {Lagendijk},\ and\
  \citenamefont {Righini}}]{Wiersma1997}%
  \BibitemOpen
  \bibfield  {author} {\bibinfo {author} {\bibfnamefont {D.~S.}\ \bibnamefont
  {Wiersma}}, \bibinfo {author} {\bibfnamefont {P.}~\bibnamefont {Bartolini}},
  \bibinfo {author} {\bibfnamefont {A.}~\bibnamefont {Lagendijk}}, \ and\
  \bibinfo {author} {\bibfnamefont {R.}~\bibnamefont {Righini}},\ }\href
  {\doibase 10.1038/37757} {\bibfield  {journal} {\bibinfo  {journal} {Nature}\
  }\textbf {\bibinfo {volume} {390}},\ \bibinfo {pages} {671} (\bibinfo {year}
  {1997})}\BibitemShut {NoStop}%
\bibitem [{\citenamefont {Rose}(2013)}]{Rose2013}%
  \BibitemOpen
  \bibfield  {author} {\bibinfo {author} {\bibfnamefont {H.}~\bibnamefont
  {Rose}},\ }\href {https://books.google.fr/books?id=rhy6BQAAQBAJ} {\emph
  {\bibinfo {title} {{Geometrical Charged-Particle Optics}}}},\ Springer Series
  in Optical Sciences\ (\bibinfo  {publisher} {Springer Berlin Heidelberg},\
  \bibinfo {year} {2013})\BibitemShut {NoStop}%
\bibitem [{\citenamefont {Scherzer}(1949)}]{Scherzer1949}%
  \BibitemOpen
  \bibfield  {author} {\bibinfo {author} {\bibfnamefont {O.}~\bibnamefont
  {Scherzer}},\ }\href@noop {} {\bibfield  {journal} {\bibinfo  {journal}
  {Journal of Applied Physics}\ }\textbf {\bibinfo {volume} {20}} (\bibinfo
  {year} {1949})}\BibitemShut {NoStop}%
\bibitem [{\citenamefont {Haider}\ \emph {et~al.}(1998)\citenamefont {Haider},
  \citenamefont {Uhlemann}, \citenamefont {Schwan}, \citenamefont {Rose},
  \citenamefont {Kabius},\ and\ \citenamefont {Urban}}]{Haider1998}%
  \BibitemOpen
  \bibfield  {author} {\bibinfo {author} {\bibfnamefont {M.}~\bibnamefont
  {Haider}}, \bibinfo {author} {\bibfnamefont {S.}~\bibnamefont {Uhlemann}},
  \bibinfo {author} {\bibfnamefont {E.}~\bibnamefont {Schwan}}, \bibinfo
  {author} {\bibfnamefont {G.}~\bibnamefont {Rose}}, \bibinfo {author}
  {\bibfnamefont {B.}~\bibnamefont {Kabius}}, \ and\ \bibinfo {author}
  {\bibfnamefont {K.}~\bibnamefont {Urban}},\ }\href {\doibase 10.1038/33823}
  {\bibfield  {journal} {\bibinfo  {journal} {Nature}\ }\textbf {\bibinfo
  {volume} {392}},\ \bibinfo {pages} {768} (\bibinfo {year}
  {1998})}\BibitemShut {NoStop}%
\bibitem [{\citenamefont {Gabor}(1948)}]{Gabor1948}%
  \BibitemOpen
  \bibfield  {author} {\bibinfo {author} {\bibfnamefont {D.}~\bibnamefont
  {Gabor}},\ }\href {\doibase 10.1038/162680a0} {\bibfield  {journal} {\bibinfo
   {journal} {Nature}\ }\textbf {\bibinfo {volume} {161}},\ \bibinfo {pages}
  {777} (\bibinfo {year} {1948})}\BibitemShut {NoStop}%
\bibitem [{\citenamefont {Bliokh}\ \emph {et~al.}(2007)\citenamefont {Bliokh},
  \citenamefont {Bliokh}, \citenamefont {Savel'ev},\ and\ \citenamefont
  {Nori}}]{Bliokh2007}%
  \BibitemOpen
  \bibfield  {author} {\bibinfo {author} {\bibfnamefont {K.~Y.}\ \bibnamefont
  {Bliokh}}, \bibinfo {author} {\bibfnamefont {Y.~P.}\ \bibnamefont {Bliokh}},
  \bibinfo {author} {\bibfnamefont {S.}~\bibnamefont {Savel'ev}}, \ and\
  \bibinfo {author} {\bibfnamefont {F.}~\bibnamefont {Nori}},\ }\href {\doibase
  10.1103/PhysRevLett.99.190404} {\bibfield  {journal} {\bibinfo  {journal}
  {Phys. Rev. Lett.}\ }\textbf {\bibinfo {volume} {99}},\ \bibinfo {pages}
  {190404} (\bibinfo {year} {2007})}\BibitemShut {NoStop}%
\bibitem [{\citenamefont {Uchida}\ and\ \citenamefont
  {Tonomura}(2010)}]{Uchida2010}%
  \BibitemOpen
  \bibfield  {author} {\bibinfo {author} {\bibfnamefont {M.}~\bibnamefont
  {Uchida}}\ and\ \bibinfo {author} {\bibfnamefont {A.}~\bibnamefont
  {Tonomura}},\ }\href {\doibase 10.1038/nature08904} {\bibfield  {journal}
  {\bibinfo  {journal} {Nature}\ }\textbf {\bibinfo {volume} {464}},\ \bibinfo
  {pages} {737} (\bibinfo {year} {2010})}\BibitemShut {NoStop}%
\bibitem [{\citenamefont {Verbeeck}\ \emph {et~al.}(2010)\citenamefont
  {Verbeeck}, \citenamefont {Tian},\ and\ \citenamefont
  {Schattschneider}}]{Verbeeck2010}%
  \BibitemOpen
  \bibfield  {author} {\bibinfo {author} {\bibfnamefont {J.}~\bibnamefont
  {Verbeeck}}, \bibinfo {author} {\bibfnamefont {H.}~\bibnamefont {Tian}}, \
  and\ \bibinfo {author} {\bibfnamefont {P.}~\bibnamefont {Schattschneider}},\
  }\href {\doibase 10.1038/nature09366} {\bibfield  {journal} {\bibinfo
  {journal} {Nature}\ }\textbf {\bibinfo {volume} {467}},\ \bibinfo {pages}
  {301} (\bibinfo {year} {2010})}\BibitemShut {NoStop}%
\bibitem [{\citenamefont {McMorran}\ \emph {et~al.}(2011)\citenamefont
  {McMorran}, \citenamefont {Agrawal}, \citenamefont {Anderson}, \citenamefont
  {Herzing}, \citenamefont {Lezec}, \citenamefont {McClelland},\ and\
  \citenamefont {Unguris}}]{Mcmorran2011}%
  \BibitemOpen
  \bibfield  {author} {\bibinfo {author} {\bibfnamefont {B.~J.}\ \bibnamefont
  {McMorran}}, \bibinfo {author} {\bibfnamefont {A.}~\bibnamefont {Agrawal}},
  \bibinfo {author} {\bibfnamefont {I.~M.}\ \bibnamefont {Anderson}}, \bibinfo
  {author} {\bibfnamefont {A.~A.}\ \bibnamefont {Herzing}}, \bibinfo {author}
  {\bibfnamefont {H.~J.}\ \bibnamefont {Lezec}}, \bibinfo {author}
  {\bibfnamefont {J.~J.}\ \bibnamefont {McClelland}}, \ and\ \bibinfo {author}
  {\bibfnamefont {J.}~\bibnamefont {Unguris}},\ }\href {\doibase
  10.1126/science.1198804} {\bibfield  {journal} {\bibinfo  {journal}
  {Science}\ }\textbf {\bibinfo {volume} {331}},\ \bibinfo {pages} {192}
  (\bibinfo {year} {2011})}\BibitemShut {NoStop}%
\bibitem [{\citenamefont {Nye}\ and\ \citenamefont {Berry}(1974)}]{Nye1974}%
  \BibitemOpen
  \bibfield  {author} {\bibinfo {author} {\bibfnamefont {J.~F.}\ \bibnamefont
  {Nye}}\ and\ \bibinfo {author} {\bibfnamefont {M.~V.}\ \bibnamefont
  {Berry}},\ }\href {\doibase 10.1098/rspa.1974.0012} {\bibfield  {journal}
  {\bibinfo  {journal} {Proceedings of the Royal Society A: Mathematical,
  Physical and Engineering Sciences}\ }\textbf {\bibinfo {volume} {336}},\
  \bibinfo {pages} {165} (\bibinfo {year} {1974})}\BibitemShut {NoStop}%
\bibitem [{\citenamefont {Allen}\ \emph {et~al.}(1992)\citenamefont {Allen},
  \citenamefont {Beijersbergen}, \citenamefont {Spreeuw},\ and\ \citenamefont
  {Woerdman}}]{Allen1992}%
  \BibitemOpen
  \bibfield  {author} {\bibinfo {author} {\bibfnamefont {L.}~\bibnamefont
  {Allen}}, \bibinfo {author} {\bibfnamefont {M.~W.}\ \bibnamefont
  {Beijersbergen}}, \bibinfo {author} {\bibfnamefont {R.~J.~C.}\ \bibnamefont
  {Spreeuw}}, \ and\ \bibinfo {author} {\bibfnamefont {J.~P.}\ \bibnamefont
  {Woerdman}},\ }\href {\doibase 10.1103/PhysRevA.45.8185} {\bibfield
  {journal} {\bibinfo  {journal} {Phys. Rev. A}\ }\textbf {\bibinfo {volume}
  {45}},\ \bibinfo {pages} {8185} (\bibinfo {year} {1992})}\BibitemShut
  {NoStop}%
\bibitem [{\citenamefont {{De Garc{\'{i}}a Abajo}}\ \emph
  {et~al.}(2009)\citenamefont {{De Garc{\'{i}}a Abajo}}, \citenamefont
  {Estrada},\ and\ \citenamefont {Meseguer}}]{DeGarciaAbajo2009a}%
  \BibitemOpen
  \bibfield  {author} {\bibinfo {author} {\bibfnamefont {F.~J.}\ \bibnamefont
  {{De Garc{\'{i}}a Abajo}}}, \bibinfo {author} {\bibfnamefont
  {H.}~\bibnamefont {Estrada}}, \ and\ \bibinfo {author} {\bibfnamefont
  {F.}~\bibnamefont {Meseguer}},\ }\href {\doibase
  10.1088/1367-2630/11/9/093013} {\bibfield  {journal} {\bibinfo  {journal}
  {New Journal of Physics}\ }\textbf {\bibinfo {volume} {11}},\ \bibinfo
  {pages} {1} (\bibinfo {year} {2009})}\BibitemShut {NoStop}%
\bibitem [{\citenamefont {Losquin}\ \emph {et~al.}(2015)\citenamefont
  {Losquin}, \citenamefont {Zagonel}, \citenamefont {Myroshnychenko},
  \citenamefont {Rodr{\'{i}}guez-Gonz{\'{a}}lez}, \citenamefont {Tenc{\'{e}}},
  \citenamefont {Scarabelli}, \citenamefont {F{\"{o}}rstner}, \citenamefont
  {Liz-Marz{\'{a}}n}, \citenamefont {{Garc{\'{i}}a De Abajo}}, \citenamefont
  {St{\'{e}}phan},\ and\ \citenamefont {Kociak}}]{Losquin2015}%
  \BibitemOpen
  \bibfield  {author} {\bibinfo {author} {\bibfnamefont {A.}~\bibnamefont
  {Losquin}}, \bibinfo {author} {\bibfnamefont {L.~F.}\ \bibnamefont
  {Zagonel}}, \bibinfo {author} {\bibfnamefont {V.}~\bibnamefont
  {Myroshnychenko}}, \bibinfo {author} {\bibfnamefont {B.}~\bibnamefont
  {Rodr{\'{i}}guez-Gonz{\'{a}}lez}}, \bibinfo {author} {\bibfnamefont
  {M.}~\bibnamefont {Tenc{\'{e}}}}, \bibinfo {author} {\bibfnamefont
  {L.}~\bibnamefont {Scarabelli}}, \bibinfo {author} {\bibfnamefont
  {J.}~\bibnamefont {F{\"{o}}rstner}}, \bibinfo {author} {\bibfnamefont
  {L.~M.}\ \bibnamefont {Liz-Marz{\'{a}}n}}, \bibinfo {author} {\bibfnamefont
  {F.~J.}\ \bibnamefont {{Garc{\'{i}}a De Abajo}}}, \bibinfo {author}
  {\bibfnamefont {O.}~\bibnamefont {St{\'{e}}phan}}, \ and\ \bibinfo {author}
  {\bibfnamefont {M.}~\bibnamefont {Kociak}},\ }\href {\doibase
  10.1021/nl5043775} {\bibfield  {journal} {\bibinfo  {journal} {Nano Letters}\
  }\textbf {\bibinfo {volume} {15}},\ \bibinfo {pages} {1229} (\bibinfo {year}
  {2015})}\BibitemShut {NoStop}%
\bibitem [{\citenamefont {{Garc{\'{i}}a de Abajo}}\ and\ \citenamefont
  {Kociak}(2008)}]{GarciadeAbajo2008}%
  \BibitemOpen
  \bibfield  {author} {\bibinfo {author} {\bibfnamefont {F.~J.}\ \bibnamefont
  {{Garc{\'{i}}a de Abajo}}}\ and\ \bibinfo {author} {\bibfnamefont
  {M.}~\bibnamefont {Kociak}},\ }\href {\doibase
  10.1103/PhysRevLett.100.106804} {\bibfield  {journal} {\bibinfo  {journal}
  {Physical Review Letters}\ }\textbf {\bibinfo {volume} {100}},\ \bibinfo
  {pages} {106804} (\bibinfo {year} {2008})}\BibitemShut {NoStop}%
\bibitem [{\citenamefont {Losquin}\ and\ \citenamefont
  {Kociak}(2015)}]{Losquin2015b}%
  \BibitemOpen
  \bibfield  {author} {\bibinfo {author} {\bibfnamefont {A.}~\bibnamefont
  {Losquin}}\ and\ \bibinfo {author} {\bibfnamefont {M.}~\bibnamefont
  {Kociak}},\ }\href {\doibase 10.1021/acsphotonics.5b00416} {\bibfield
  {journal} {\bibinfo  {journal} {ACS Photonics}\ }\textbf {\bibinfo {volume}
  {2}},\ \bibinfo {pages} {1619} (\bibinfo {year} {2015})}\BibitemShut
  {NoStop}%
\bibitem [{\citenamefont {Kim}\ and\ \citenamefont {Gbur}(2012)}]{Kim2012}%
  \BibitemOpen
  \bibfield  {author} {\bibinfo {author} {\bibfnamefont {S.~M.}\ \bibnamefont
  {Kim}}\ and\ \bibinfo {author} {\bibfnamefont {G.}~\bibnamefont {Gbur}},\
  }\href {\doibase 10.1103/PhysRevA.86.043814} {\bibfield  {journal} {\bibinfo
  {journal} {Phys. Rev. A}\ }\textbf {\bibinfo {volume} {86}},\ \bibinfo
  {pages} {043814} (\bibinfo {year} {2012})}\BibitemShut {NoStop}%
\bibitem [{\citenamefont {Collins}\ \emph {et~al.}(2017)\citenamefont
  {Collins}, \citenamefont {Kuppe}, \citenamefont {Hooper}, \citenamefont
  {Sibilia}, \citenamefont {Centini},\ and\ \citenamefont
  {Valev}}]{Collins2017}%
  \BibitemOpen
  \bibfield  {author} {\bibinfo {author} {\bibfnamefont {J.~T.}\ \bibnamefont
  {Collins}}, \bibinfo {author} {\bibfnamefont {C.}~\bibnamefont {Kuppe}},
  \bibinfo {author} {\bibfnamefont {D.~C.}\ \bibnamefont {Hooper}}, \bibinfo
  {author} {\bibfnamefont {C.}~\bibnamefont {Sibilia}}, \bibinfo {author}
  {\bibfnamefont {M.}~\bibnamefont {Centini}}, \ and\ \bibinfo {author}
  {\bibfnamefont {V.~K.}\ \bibnamefont {Valev}},\ }\href {\doibase
  10.1002/adom.201700182} {\bibfield  {journal} {\bibinfo  {journal} {Advanced
  Optical Materials}\ }\textbf {\bibinfo {volume} {5}},\ \bibinfo {pages}
  {1700182} (\bibinfo {year} {2017})}\BibitemShut {NoStop}%
\bibitem [{\citenamefont {Tullius}\ \emph {et~al.}(2015)\citenamefont
  {Tullius}, \citenamefont {Karimullah}, \citenamefont {Rodier}, \citenamefont
  {Fitzpatrick}, \citenamefont {Gadegaard}, \citenamefont {Barron},
  \citenamefont {Rotello}, \citenamefont {Cooke}, \citenamefont {Lapthorn},\
  and\ \citenamefont {Kadodwala}}]{Tullius2015}%
  \BibitemOpen
  \bibfield  {author} {\bibinfo {author} {\bibfnamefont {R.}~\bibnamefont
  {Tullius}}, \bibinfo {author} {\bibfnamefont {A.~S.}\ \bibnamefont
  {Karimullah}}, \bibinfo {author} {\bibfnamefont {M.}~\bibnamefont {Rodier}},
  \bibinfo {author} {\bibfnamefont {B.}~\bibnamefont {Fitzpatrick}}, \bibinfo
  {author} {\bibfnamefont {N.}~\bibnamefont {Gadegaard}}, \bibinfo {author}
  {\bibfnamefont {L.~D.}\ \bibnamefont {Barron}}, \bibinfo {author}
  {\bibfnamefont {V.~M.}\ \bibnamefont {Rotello}}, \bibinfo {author}
  {\bibfnamefont {G.}~\bibnamefont {Cooke}}, \bibinfo {author} {\bibfnamefont
  {A.}~\bibnamefont {Lapthorn}}, \ and\ \bibinfo {author} {\bibfnamefont
  {M.}~\bibnamefont {Kadodwala}},\ }\href {\doibase 10.1021/jacs.5b04806}
  {\bibfield  {journal} {\bibinfo  {journal} {Journal of the American Chemical
  Society}\ }\textbf {\bibinfo {volume} {137}},\ \bibinfo {pages} {8380}
  (\bibinfo {year} {2015})}\BibitemShut {NoStop}%
\bibitem [{\citenamefont {Sch\"aferling}\ \emph {et~al.}(2012)\citenamefont
  {Sch\"aferling}, \citenamefont {Dregely}, \citenamefont {Hentschel},\ and\
  \citenamefont {Giessen}}]{Schaeferling2012}%
  \BibitemOpen
  \bibfield  {author} {\bibinfo {author} {\bibfnamefont {M.}~\bibnamefont
  {Sch\"aferling}}, \bibinfo {author} {\bibfnamefont {D.}~\bibnamefont
  {Dregely}}, \bibinfo {author} {\bibfnamefont {M.}~\bibnamefont {Hentschel}},
  \ and\ \bibinfo {author} {\bibfnamefont {H.}~\bibnamefont {Giessen}},\ }\href
  {\doibase 10.1103/PhysRevX.2.031010} {\bibfield  {journal} {\bibinfo
  {journal} {Phys. Rev. X}\ }\textbf {\bibinfo {volume} {2}},\ \bibinfo {pages}
  {031010} (\bibinfo {year} {2012})}\BibitemShut {NoStop}%
\bibitem [{\citenamefont {Asenjo-Garcia}\ and\ \citenamefont {{Garc{\'{i}}a De
  Abajo}}(2014)}]{Asenjo-Garcia2014}%
  \BibitemOpen
  \bibfield  {author} {\bibinfo {author} {\bibfnamefont {A.}~\bibnamefont
  {Asenjo-Garcia}}\ and\ \bibinfo {author} {\bibfnamefont {F.~J.}\ \bibnamefont
  {{Garc{\'{i}}a De Abajo}}},\ }\href {\doibase 10.1103/PhysRevLett.113.066102}
  {\bibfield  {journal} {\bibinfo  {journal} {Physical Review Letters}\
  }\textbf {\bibinfo {volume} {113}},\ \bibinfo {pages} {1} (\bibinfo {year}
  {2014})}\BibitemShut {NoStop}%
\bibitem [{\citenamefont {Guzzinati}\ \emph {et~al.}(2017)\citenamefont
  {Guzzinati}, \citenamefont {B{\'{e}}ch{\'{e}}}, \citenamefont
  {Louren{\c{c}}o-Martins}, \citenamefont {Martin}, \citenamefont {Kociak},\
  and\ \citenamefont {Verbeeck}}]{Guzzinati2017}%
  \BibitemOpen
  \bibfield  {author} {\bibinfo {author} {\bibfnamefont {G.}~\bibnamefont
  {Guzzinati}}, \bibinfo {author} {\bibfnamefont {A.}~\bibnamefont
  {B{\'{e}}ch{\'{e}}}}, \bibinfo {author} {\bibfnamefont {H.}~\bibnamefont
  {Louren{\c{c}}o-Martins}}, \bibinfo {author} {\bibfnamefont {J.}~\bibnamefont
  {Martin}}, \bibinfo {author} {\bibfnamefont {M.}~\bibnamefont {Kociak}}, \
  and\ \bibinfo {author} {\bibfnamefont {J.}~\bibnamefont {Verbeeck}},\ }\href
  {\doibase 10.1038/ncomms14999} {\bibfield  {journal} {\bibinfo  {journal}
  {Nature Communications}\ }\textbf {\bibinfo {volume} {8}},\ \bibinfo {pages}
  {14999} (\bibinfo {year} {2017})}\BibitemShut {NoStop}%
\bibitem [{\citenamefont {Louren{\c{c}}o-Martins}\ \emph
  {et~al.}(2020)\citenamefont {Louren{\c{c}}o-Martins}, \citenamefont {Lubk},\
  and\ \citenamefont {Kociak}}]{Lourenco-Martins2020a}%
  \BibitemOpen
  \bibfield  {author} {\bibinfo {author} {\bibfnamefont {H.}~\bibnamefont
  {Louren{\c{c}}o-Martins}}, \bibinfo {author} {\bibfnamefont {A.}~\bibnamefont
  {Lubk}}, \ and\ \bibinfo {author} {\bibfnamefont {M.}~\bibnamefont
  {Kociak}},\ }\href@noop {} {\bibfield  {journal} {\bibinfo  {journal} {in
  preparation}\ } (\bibinfo {year} {2020})}\BibitemShut {NoStop}%
\bibitem [{\citenamefont {Schattschneider}\ \emph {et~al.}(2014)\citenamefont
  {Schattschneider}, \citenamefont {Löffler}, \citenamefont
  {Stöger-Pollach},\ and\ \citenamefont {Verbeeck}}]{Schattschneider2014}%
  \BibitemOpen
  \bibfield  {author} {\bibinfo {author} {\bibfnamefont {P.}~\bibnamefont
  {Schattschneider}}, \bibinfo {author} {\bibfnamefont {S.}~\bibnamefont
  {Löffler}}, \bibinfo {author} {\bibfnamefont {M.}~\bibnamefont
  {Stöger-Pollach}}, \ and\ \bibinfo {author} {\bibfnamefont {J.}~\bibnamefont
  {Verbeeck}},\ }\href {\doibase
  https://doi.org/10.1016/j.ultramic.2013.07.012} {\bibfield  {journal}
  {\bibinfo  {journal} {Ultramicroscopy}\ }\textbf {\bibinfo {volume} {136}},\
  \bibinfo {pages} {81 } (\bibinfo {year} {2014})}\BibitemShut {NoStop}%
\bibitem [{\citenamefont {Ugarte}\ and\ \citenamefont
  {Ducati}(2016)}]{Ugarte2016}%
  \BibitemOpen
  \bibfield  {author} {\bibinfo {author} {\bibfnamefont {D.}~\bibnamefont
  {Ugarte}}\ and\ \bibinfo {author} {\bibfnamefont {C.}~\bibnamefont
  {Ducati}},\ }\href {\doibase 10.1103/PhysRevB.93.205418} {\bibfield
  {journal} {\bibinfo  {journal} {Physical Review B}\ }\textbf {\bibinfo
  {volume} {93}},\ \bibinfo {pages} {1} (\bibinfo {year} {2016})}\BibitemShut
  {NoStop}%
\bibitem [{\citenamefont {Zanfrognini}\ \emph {et~al.}(2019)\citenamefont
  {Zanfrognini}, \citenamefont {Rotunno}, \citenamefont {Frabboni},
  \citenamefont {Sit}, \citenamefont {Karimi}, \citenamefont {Hohenester},\
  and\ \citenamefont {Grillo}}]{Zanfrognini2019}%
  \BibitemOpen
  \bibfield  {author} {\bibinfo {author} {\bibfnamefont {M.}~\bibnamefont
  {Zanfrognini}}, \bibinfo {author} {\bibfnamefont {E.}~\bibnamefont
  {Rotunno}}, \bibinfo {author} {\bibfnamefont {S.}~\bibnamefont {Frabboni}},
  \bibinfo {author} {\bibfnamefont {A.}~\bibnamefont {Sit}}, \bibinfo {author}
  {\bibfnamefont {E.}~\bibnamefont {Karimi}}, \bibinfo {author} {\bibfnamefont
  {U.}~\bibnamefont {Hohenester}}, \ and\ \bibinfo {author} {\bibfnamefont
  {V.}~\bibnamefont {Grillo}},\ }\href {\doibase 10.1021/acsphotonics.9b00131}
  {\bibfield  {journal} {\bibinfo  {journal} {ACS Photonics}\ }\textbf
  {\bibinfo {volume} {6}},\ \bibinfo {pages} {620} (\bibinfo {year}
  {2019})}\BibitemShut {NoStop}%
\bibitem [{\citenamefont {Ouyang}\ and\ \citenamefont
  {Isaacson}(1989)}]{Ouyang1989}%
  \BibitemOpen
  \bibfield  {author} {\bibinfo {author} {\bibfnamefont {F.}~\bibnamefont
  {Ouyang}}\ and\ \bibinfo {author} {\bibfnamefont {M.}~\bibnamefont
  {Isaacson}},\ }\href {\doibase 10.1080/13642818908205921} {\bibfield
  {journal} {\bibinfo  {journal} {Philosophical Magazine Part B}\ }\textbf
  {\bibinfo {volume} {60}},\ \bibinfo {pages} {481} (\bibinfo {year}
  {1989})}\BibitemShut {NoStop}%
\bibitem [{\citenamefont {Boudarham}\ and\ \citenamefont
  {Kociak}(2012)}]{Boudarham2012}%
  \BibitemOpen
  \bibfield  {author} {\bibinfo {author} {\bibfnamefont {G.}~\bibnamefont
  {Boudarham}}\ and\ \bibinfo {author} {\bibfnamefont {M.}~\bibnamefont
  {Kociak}},\ }\href {\doibase 10.1103/PhysRevB.85.245447} {\bibfield
  {journal} {\bibinfo  {journal} {Phys. Rev. B}\ }\textbf {\bibinfo {volume}
  {85}},\ \bibinfo {pages} {245447} (\bibinfo {year} {2012})}\BibitemShut
  {NoStop}%
\bibitem [{\citenamefont {Hohenester}(2014)}]{Hohenester2014}%
  \BibitemOpen
  \bibfield  {author} {\bibinfo {author} {\bibfnamefont {U.}~\bibnamefont
  {Hohenester}},\ }\href {\doibase https://doi.org/10.1016/j.cpc.2013.12.010}
  {\bibfield  {journal} {\bibinfo  {journal} {Computer Physics Communications}\
  }\textbf {\bibinfo {volume} {185}},\ \bibinfo {pages} {1177 } (\bibinfo
  {year} {2014})}\BibitemShut {NoStop}%
\bibitem [{\citenamefont {{Garc{\'{i}}a de Abajo}}(2010)}]{GarciadeAbajo2010}%
  \BibitemOpen
  \bibfield  {author} {\bibinfo {author} {\bibfnamefont {F.~J.}\ \bibnamefont
  {{Garc{\'{i}}a de Abajo}}},\ }\href {\doibase 10.1103/RevModPhys.82.209}
  {\bibfield  {journal} {\bibinfo  {journal} {Reviews of Modern Physics}\
  }\textbf {\bibinfo {volume} {82}},\ \bibinfo {pages} {209} (\bibinfo {year}
  {2010})}\BibitemShut {NoStop}%
\bibitem [{\citenamefont {Schattschneider}\ \emph {et~al.}(2012)\citenamefont
  {Schattschneider}, \citenamefont {St\"oger-Pollach},\ and\ \citenamefont
  {Verbeeck}}]{Schattschneider2012}%
  \BibitemOpen
  \bibfield  {author} {\bibinfo {author} {\bibfnamefont {P.}~\bibnamefont
  {Schattschneider}}, \bibinfo {author} {\bibfnamefont {M.}~\bibnamefont
  {St\"oger-Pollach}}, \ and\ \bibinfo {author} {\bibfnamefont
  {J.}~\bibnamefont {Verbeeck}},\ }\href {\doibase
  10.1103/PhysRevLett.109.084801} {\bibfield  {journal} {\bibinfo  {journal}
  {Phys. Rev. Lett.}\ }\textbf {\bibinfo {volume} {109}},\ \bibinfo {pages}
  {084801} (\bibinfo {year} {2012})}\BibitemShut {NoStop}%
\bibitem [{\citenamefont {Grillo}\ \emph {et~al.}(2017)\citenamefont {Grillo},
  \citenamefont {Tavabi}, \citenamefont {Venturi}, \citenamefont {Larocque},
  \citenamefont {Balboni}, \citenamefont {Gazzadi}, \citenamefont {Frabboni},
  \citenamefont {Lu}, \citenamefont {Mafakheri}, \citenamefont {Bouchard},
  \citenamefont {Dunin-Borkowski}, \citenamefont {Boyd}, \citenamefont
  {Lavery}, \citenamefont {Padgett},\ and\ \citenamefont
  {Karimi}}]{Grillo2017}%
  \BibitemOpen
  \bibfield  {author} {\bibinfo {author} {\bibfnamefont {V.}~\bibnamefont
  {Grillo}}, \bibinfo {author} {\bibfnamefont {A.~H.}\ \bibnamefont {Tavabi}},
  \bibinfo {author} {\bibfnamefont {F.}~\bibnamefont {Venturi}}, \bibinfo
  {author} {\bibfnamefont {H.}~\bibnamefont {Larocque}}, \bibinfo {author}
  {\bibfnamefont {R.}~\bibnamefont {Balboni}}, \bibinfo {author} {\bibfnamefont
  {G.~C.}\ \bibnamefont {Gazzadi}}, \bibinfo {author} {\bibfnamefont
  {S.}~\bibnamefont {Frabboni}}, \bibinfo {author} {\bibfnamefont {P.~H.}\
  \bibnamefont {Lu}}, \bibinfo {author} {\bibfnamefont {E.}~\bibnamefont
  {Mafakheri}}, \bibinfo {author} {\bibfnamefont {F.}~\bibnamefont {Bouchard}},
  \bibinfo {author} {\bibfnamefont {R.~E.}\ \bibnamefont {Dunin-Borkowski}},
  \bibinfo {author} {\bibfnamefont {R.~W.}\ \bibnamefont {Boyd}}, \bibinfo
  {author} {\bibfnamefont {M.~P.}\ \bibnamefont {Lavery}}, \bibinfo {author}
  {\bibfnamefont {M.~J.}\ \bibnamefont {Padgett}}, \ and\ \bibinfo {author}
  {\bibfnamefont {E.}~\bibnamefont {Karimi}},\ }\href {\doibase
  10.1038/ncomms15536} {\bibfield  {journal} {\bibinfo  {journal} {Nature
  Communications}\ }\textbf {\bibinfo {volume} {8}},\ \bibinfo {pages} {6}
  (\bibinfo {year} {2017})}\BibitemShut {NoStop}%
\bibitem [{\citenamefont {Pozzi}\ \emph {et~al.}(2020)\citenamefont {Pozzi},
  \citenamefont {Grillo}, \citenamefont {Lu}, \citenamefont {Tavabi},
  \citenamefont {Karimi},\ and\ \citenamefont {Dunin-Borkowski}}]{Pozzi2020}%
  \BibitemOpen
  \bibfield  {author} {\bibinfo {author} {\bibfnamefont {G.}~\bibnamefont
  {Pozzi}}, \bibinfo {author} {\bibfnamefont {V.}~\bibnamefont {Grillo}},
  \bibinfo {author} {\bibfnamefont {P.-H.}\ \bibnamefont {Lu}}, \bibinfo
  {author} {\bibfnamefont {A.~H.}\ \bibnamefont {Tavabi}}, \bibinfo {author}
  {\bibfnamefont {E.}~\bibnamefont {Karimi}}, \ and\ \bibinfo {author}
  {\bibfnamefont {R.~E.}\ \bibnamefont {Dunin-Borkowski}},\ }\href {\doibase
  https://doi.org/10.1016/j.ultramic.2019.112861} {\bibfield  {journal}
  {\bibinfo  {journal} {Ultramicroscopy}\ }\textbf {\bibinfo {volume} {208}},\
  \bibinfo {pages} {112861} (\bibinfo {year} {2020})}\BibitemShut {NoStop}%
\bibitem [{\citenamefont {Yin}\ \emph {et~al.}(2013)\citenamefont {Yin},
  \citenamefont {Sch{\"{a}}ferling}, \citenamefont {Metzger},\ and\
  \citenamefont {Giessen}}]{Yin2013a}%
  \BibitemOpen
  \bibfield  {author} {\bibinfo {author} {\bibfnamefont {X.}~\bibnamefont
  {Yin}}, \bibinfo {author} {\bibfnamefont {M.}~\bibnamefont
  {Sch{\"{a}}ferling}}, \bibinfo {author} {\bibfnamefont {B.}~\bibnamefont
  {Metzger}}, \ and\ \bibinfo {author} {\bibfnamefont {H.}~\bibnamefont
  {Giessen}},\ }\href {\doibase 10.1021/nl403705k} {\bibfield  {journal}
  {\bibinfo  {journal} {Nano Letters}\ }\textbf {\bibinfo {volume} {13}},\
  \bibinfo {pages} {6238} (\bibinfo {year} {2013})}\BibitemShut {NoStop}%
\bibitem [{\citenamefont {Lu}\ \emph {et~al.}(2014)\citenamefont {Lu},
  \citenamefont {Wu}, \citenamefont {Zhu}, \citenamefont {Zhao}, \citenamefont
  {Wang}, \citenamefont {Zhan},\ and\ \citenamefont {Ni}}]{Lu2014}%
  \BibitemOpen
  \bibfield  {author} {\bibinfo {author} {\bibfnamefont {X.}~\bibnamefont
  {Lu}}, \bibinfo {author} {\bibfnamefont {J.}~\bibnamefont {Wu}}, \bibinfo
  {author} {\bibfnamefont {Q.}~\bibnamefont {Zhu}}, \bibinfo {author}
  {\bibfnamefont {J.}~\bibnamefont {Zhao}}, \bibinfo {author} {\bibfnamefont
  {Q.}~\bibnamefont {Wang}}, \bibinfo {author} {\bibfnamefont {L.}~\bibnamefont
  {Zhan}}, \ and\ \bibinfo {author} {\bibfnamefont {W.}~\bibnamefont {Ni}},\
  }\href {\doibase 10.1039/c4nr04433a} {\bibfield  {journal} {\bibinfo
  {journal} {Nanoscale}\ }\textbf {\bibinfo {volume} {6}},\ \bibinfo {pages}
  {14244} (\bibinfo {year} {2014})}\BibitemShut {NoStop}%
\bibitem [{\citenamefont {{R. Harvey}}\ \emph {et~al.}(2020)\citenamefont {{R.
  Harvey}}, \citenamefont {Henke}, \citenamefont {Kfir}, \citenamefont
  {Louren{\c{c}}o-Martins}, \citenamefont {Feist}, \citenamefont {{Javier
  Garc{\'{i}}a de Abajo}},\ and\ \citenamefont {Ropers}}]{Harvey2020}%
  \BibitemOpen
  \bibfield  {author} {\bibinfo {author} {\bibfnamefont {T.}~\bibnamefont {{R.
  Harvey}}}, \bibinfo {author} {\bibfnamefont {J.-W.}\ \bibnamefont {Henke}},
  \bibinfo {author} {\bibfnamefont {O.}~\bibnamefont {Kfir}}, \bibinfo {author}
  {\bibfnamefont {H.}~\bibnamefont {Louren{\c{c}}o-Martins}}, \bibinfo {author}
  {\bibfnamefont {A.}~\bibnamefont {Feist}}, \bibinfo {author} {\bibfnamefont
  {F.}~\bibnamefont {{Javier Garc{\'{i}}a de Abajo}}}, \ and\ \bibinfo {author}
  {\bibfnamefont {C.}~\bibnamefont {Ropers}},\ }\href {\doibase
  10.1021/acs.nanolett.0c01130} {\bibfield  {journal} {\bibinfo  {journal}
  {Nano Letters}\ }\textbf {\bibinfo {volume} {0}} (\bibinfo {year} {2020}),\
  10.1021/acs.nanolett.0c01130}\BibitemShut {NoStop}%
\bibitem [{\citenamefont {Pham}\ \emph {et~al.}(2016)\citenamefont {Pham},
  \citenamefont {Berthel}, \citenamefont {Jiang}, \citenamefont {Bellessa},
  \citenamefont {Huant}, \citenamefont {Genet},\ and\ \citenamefont
  {Drezet}}]{Pham2016}%
  \BibitemOpen
  \bibfield  {author} {\bibinfo {author} {\bibfnamefont {A.}~\bibnamefont
  {Pham}}, \bibinfo {author} {\bibfnamefont {M.}~\bibnamefont {Berthel}},
  \bibinfo {author} {\bibfnamefont {Q.}~\bibnamefont {Jiang}}, \bibinfo
  {author} {\bibfnamefont {J.}~\bibnamefont {Bellessa}}, \bibinfo {author}
  {\bibfnamefont {S.}~\bibnamefont {Huant}}, \bibinfo {author} {\bibfnamefont
  {C.}~\bibnamefont {Genet}}, \ and\ \bibinfo {author} {\bibfnamefont
  {A.}~\bibnamefont {Drezet}},\ }\href {\doibase 10.1103/PhysRevA.94.053850}
  {\bibfield  {journal} {\bibinfo  {journal} {Phys. Rev. A}\ }\textbf {\bibinfo
  {volume} {94}},\ \bibinfo {pages} {053850} (\bibinfo {year}
  {2016})}\BibitemShut {NoStop}%
\bibitem [{\citenamefont {Pham}\ \emph {et~al.}(2018)\citenamefont {Pham},
  \citenamefont {Zhao}, \citenamefont {Genet},\ and\ \citenamefont
  {Drezet}}]{Pham2018}%
  \BibitemOpen
  \bibfield  {author} {\bibinfo {author} {\bibfnamefont {A.}~\bibnamefont
  {Pham}}, \bibinfo {author} {\bibfnamefont {A.}~\bibnamefont {Zhao}}, \bibinfo
  {author} {\bibfnamefont {C.}~\bibnamefont {Genet}}, \ and\ \bibinfo {author}
  {\bibfnamefont {A.}~\bibnamefont {Drezet}},\ }\href {\doibase
  10.1103/PhysRevA.98.013837} {\bibfield  {journal} {\bibinfo  {journal}
  {Physical Review A}\ }\textbf {\bibinfo {volume} {98}},\ \bibinfo {pages} {1}
  (\bibinfo {year} {2018})}\BibitemShut {NoStop}%
\bibitem [{\citenamefont {Novotny}\ and\ \citenamefont
  {Hecht}(2006)}]{Novotny2006}%
  \BibitemOpen
  \bibfield  {author} {\bibinfo {author} {\bibfnamefont {L.}~\bibnamefont
  {Novotny}}\ and\ \bibinfo {author} {\bibfnamefont {B.}~\bibnamefont
  {Hecht}},\ }\href@noop {} {\emph {\bibinfo {title} {{Principles of
  Nano-Optics}}}},\ \bibinfo {edition} {cambridge}\ ed.\ (\bibinfo  {publisher}
  {Cambridge University Press;},\ \bibinfo {year} {2006})\BibitemShut {NoStop}%
\bibitem [{\citenamefont {Krehl}\ \emph {et~al.}(2018)\citenamefont {Krehl},
  \citenamefont {Guzzinati}, \citenamefont {Schultz}, \citenamefont {Potapov},
  \citenamefont {Pohl}, \citenamefont {Martin}, \citenamefont {Verbeeck},
  \citenamefont {Fery}, \citenamefont {B{\"{u}}chner},\ and\ \citenamefont
  {Lubk}}]{Krehl2018}%
  \BibitemOpen
  \bibfield  {author} {\bibinfo {author} {\bibfnamefont {J.}~\bibnamefont
  {Krehl}}, \bibinfo {author} {\bibfnamefont {G.}~\bibnamefont {Guzzinati}},
  \bibinfo {author} {\bibfnamefont {J.}~\bibnamefont {Schultz}}, \bibinfo
  {author} {\bibfnamefont {P.}~\bibnamefont {Potapov}}, \bibinfo {author}
  {\bibfnamefont {D.}~\bibnamefont {Pohl}}, \bibinfo {author} {\bibfnamefont
  {J.}~\bibnamefont {Martin}}, \bibinfo {author} {\bibfnamefont
  {J.}~\bibnamefont {Verbeeck}}, \bibinfo {author} {\bibfnamefont
  {A.}~\bibnamefont {Fery}}, \bibinfo {author} {\bibfnamefont {B.}~\bibnamefont
  {B{\"{u}}chner}}, \ and\ \bibinfo {author} {\bibfnamefont {A.}~\bibnamefont
  {Lubk}},\ }\href {\doibase 10.1038/s41467-018-06572-9} {\bibfield  {journal}
  {\bibinfo  {journal} {Nature Communications}\ }\textbf {\bibinfo {volume}
  {9}} (\bibinfo {year} {2018}),\ 10.1038/s41467-018-06572-9}\BibitemShut
  {NoStop}%
\bibitem [{\citenamefont {Burresi}\ \emph {et~al.}(2009)\citenamefont
  {Burresi}, \citenamefont {Engelen}, \citenamefont {Opheij}, \citenamefont
  {van Oosten}, \citenamefont {Mori}, \citenamefont {Baba},\ and\ \citenamefont
  {Kuipers}}]{Burresi2009}%
  \BibitemOpen
  \bibfield  {author} {\bibinfo {author} {\bibfnamefont {M.}~\bibnamefont
  {Burresi}}, \bibinfo {author} {\bibfnamefont {R.~J.~P.}\ \bibnamefont
  {Engelen}}, \bibinfo {author} {\bibfnamefont {A.}~\bibnamefont {Opheij}},
  \bibinfo {author} {\bibfnamefont {D.}~\bibnamefont {van Oosten}}, \bibinfo
  {author} {\bibfnamefont {D.}~\bibnamefont {Mori}}, \bibinfo {author}
  {\bibfnamefont {T.}~\bibnamefont {Baba}}, \ and\ \bibinfo {author}
  {\bibfnamefont {L.}~\bibnamefont {Kuipers}},\ }\href {\doibase
  10.1103/PhysRevLett.102.033902} {\bibfield  {journal} {\bibinfo  {journal}
  {Phys. Rev. Lett.}\ }\textbf {\bibinfo {volume} {102}},\ \bibinfo {pages}
  {033902} (\bibinfo {year} {2009})}\BibitemShut {NoStop}%
\bibitem [{\citenamefont {Rotenberg}\ \emph {et~al.}()\citenamefont
  {Rotenberg}, \citenamefont {le~Feber}, \citenamefont {Visser},\ and\
  \citenamefont {Kuipers}}]{Rotenberg2015}%
  \BibitemOpen
  \bibfield  {author} {\bibinfo {author} {\bibfnamefont {N.}~\bibnamefont
  {Rotenberg}}, \bibinfo {author} {\bibfnamefont {B.}~\bibnamefont {le~Feber}},
  \bibinfo {author} {\bibfnamefont {T.~D.}\ \bibnamefont {Visser}}, \ and\
  \bibinfo {author} {\bibfnamefont {L.}~\bibnamefont {Kuipers}},\ }\href@noop
  {} {\bibfield  {journal} {\bibinfo  {journal} {Optica}\ }\textbf {\bibinfo
  {volume} {2}},\ \bibinfo {pages} {540}}\BibitemShut {NoStop}%
\bibitem [{\citenamefont {Coles}\ and\ \citenamefont
  {Andrews}(2012)}]{Coles2012}%
  \BibitemOpen
  \bibfield  {author} {\bibinfo {author} {\bibfnamefont {M.~M.}\ \bibnamefont
  {Coles}}\ and\ \bibinfo {author} {\bibfnamefont {D.~L.}\ \bibnamefont
  {Andrews}},\ }\href {\doibase 10.1103/PhysRevA.85.063810} {\bibfield
  {journal} {\bibinfo  {journal} {Phys. Rev. A}\ }\textbf {\bibinfo {volume}
  {85}},\ \bibinfo {pages} {063810} (\bibinfo {year} {2012})}\BibitemShut
  {NoStop}%
\bibitem [{\citenamefont {Bliokh}\ and\ \citenamefont
  {Nori}(2011)}]{Bliokh2011}%
  \BibitemOpen
  \bibfield  {author} {\bibinfo {author} {\bibfnamefont {K.~Y.}\ \bibnamefont
  {Bliokh}}\ and\ \bibinfo {author} {\bibfnamefont {F.}~\bibnamefont {Nori}},\
  }\href {\doibase 10.1103/PhysRevA.83.021803} {\bibfield  {journal} {\bibinfo
  {journal} {Phys. Rev. A}\ }\textbf {\bibinfo {volume} {83}},\ \bibinfo
  {pages} {021803} (\bibinfo {year} {2011})}\BibitemShut {NoStop}%
\bibitem [{Note1()}]{Note1}%
  \BibitemOpen
  \bibinfo {note} {Note that, for visibility, the gap has been increased to 50
  nm on these plots}\BibitemShut {NoStop}%
\bibitem [{\citenamefont {Cameron}\ \emph {et~al.}(2012)\citenamefont
  {Cameron}, \citenamefont {Barnett},\ and\ \citenamefont
  {Yao}}]{Cameron2012b}%
  \BibitemOpen
  \bibfield  {author} {\bibinfo {author} {\bibfnamefont {R.~P.}\ \bibnamefont
  {Cameron}}, \bibinfo {author} {\bibfnamefont {S.~M.}\ \bibnamefont
  {Barnett}}, \ and\ \bibinfo {author} {\bibfnamefont {A.~M.}\ \bibnamefont
  {Yao}},\ }\href {\doibase 10.1088/1367-2630/14/5/053050} {\bibfield
  {journal} {\bibinfo  {journal} {New Journal of Physics}\ }\textbf {\bibinfo
  {volume} {14}},\ \bibinfo {pages} {053050} (\bibinfo {year}
  {2012})}\BibitemShut {NoStop}%
\bibitem [{\citenamefont {Bliokh}\ \emph {et~al.}(2013)\citenamefont {Bliokh},
  \citenamefont {Bekshaev},\ and\ \citenamefont {Nori}}]{Bliokh2013}%
  \BibitemOpen
  \bibfield  {author} {\bibinfo {author} {\bibfnamefont {K.~Y.}\ \bibnamefont
  {Bliokh}}, \bibinfo {author} {\bibfnamefont {A.~Y.}\ \bibnamefont
  {Bekshaev}}, \ and\ \bibinfo {author} {\bibfnamefont {F.}~\bibnamefont
  {Nori}},\ }\href {\doibase 10.1088/1367-2630/15/3/033026} {\bibfield
  {journal} {\bibinfo  {journal} {New Journal of Physics}\ }\textbf {\bibinfo
  {volume} {15}},\ \bibinfo {pages} {033026} (\bibinfo {year}
  {2013})}\BibitemShut {NoStop}%
\bibitem [{\citenamefont {G{\'{o}}mez-Medina}\ \emph
  {et~al.}(2008)\citenamefont {G{\'{o}}mez-Medina}, \citenamefont {Yamamoto},
  \citenamefont {Nakano},\ and\ \citenamefont {de~Abajo}}]{Gomez-Medina2008}%
  \BibitemOpen
  \bibfield  {author} {\bibinfo {author} {\bibfnamefont {R.}~\bibnamefont
  {G{\'{o}}mez-Medina}}, \bibinfo {author} {\bibfnamefont {N.}~\bibnamefont
  {Yamamoto}}, \bibinfo {author} {\bibfnamefont {M.}~\bibnamefont {Nakano}}, \
  and\ \bibinfo {author} {\bibfnamefont {F.~J.~G.}\ \bibnamefont {de~Abajo}},\
  }\href {\doibase 10.1088/1367-2630/10/10/105009} {\bibfield  {journal}
  {\bibinfo  {journal} {New Journal of Physics}\ }\textbf {\bibinfo {volume}
  {10}},\ \bibinfo {pages} {105009} (\bibinfo {year} {2008})}\BibitemShut
  {NoStop}%
\bibitem [{\citenamefont {Osorio}\ \emph {et~al.}(2016)\citenamefont {Osorio},
  \citenamefont {Coenen}, \citenamefont {Brenny}, \citenamefont {Polman},\ and\
  \citenamefont {Koenderink}}]{Osorio2016}%
  \BibitemOpen
  \bibfield  {author} {\bibinfo {author} {\bibfnamefont {C.~I.}\ \bibnamefont
  {Osorio}}, \bibinfo {author} {\bibfnamefont {T.}~\bibnamefont {Coenen}},
  \bibinfo {author} {\bibfnamefont {B.~J.}\ \bibnamefont {Brenny}}, \bibinfo
  {author} {\bibfnamefont {A.}~\bibnamefont {Polman}}, \ and\ \bibinfo {author}
  {\bibfnamefont {A.~F.}\ \bibnamefont {Koenderink}},\ }\href {\doibase
  10.1021/acsphotonics.5b00596} {\bibfield  {journal} {\bibinfo  {journal} {ACS
  Photonics}\ }\textbf {\bibinfo {volume} {3}},\ \bibinfo {pages} {147}
  (\bibinfo {year} {2016})}\BibitemShut {NoStop}%
\bibitem [{\citenamefont {Yurtsever}\ and\ \citenamefont
  {Zewail}(2012)}]{Yurtsever2012}%
  \BibitemOpen
  \bibfield  {author} {\bibinfo {author} {\bibfnamefont {A.}~\bibnamefont
  {Yurtsever}}\ and\ \bibinfo {author} {\bibfnamefont {A.~H.}\ \bibnamefont
  {Zewail}},\ }\href {\doibase 10.1021/nl301643k} {\bibfield  {journal}
  {\bibinfo  {journal} {Nano Letters}\ }\textbf {\bibinfo {volume} {12}},\
  \bibinfo {pages} {3334} (\bibinfo {year} {2012})}\BibitemShut {NoStop}%
\end{thebibliography}

\end{document}